\documentclass[12pt]{article}
 \usepackage{graphicx}
\RequirePackage{color}

 \definecolor{MyDarkGreen}{rgb}{0.02,0.60,0.06}

\title{\bf
Universal Amplitude Ratios for Constrained Critical Systems}
\author{ 
 {\it N.~Sh.~Izmailian$^{\,1,2}$} and {\it R.~Kenna$^{\,2}$,} \\~\\
$^1$ Yerevan Physics Institute, Alikhanian Brothers 2, 375036 Yerevan, \\ Armenia
{}\\~\\
$^2$ Applied Mathematics Research Centre,
Coventry University,\\
Coventry, CV1 5FB, England
{}\\~\\}
\begin{document}
%\pdfpagewidth  595pt
%\pdfpageheight 841pt

\maketitle
%-----------------------------------------------------------------------
                      {\Large
                      \begin{abstract}
%-----------------------------------------------------------------------
%
The critical properties of systems under constraint differ from their ideal counterparts through Fisher renormalization. 
The mathematical properties of Fisher renormalization applied to critical exponents are well known: 
the renormalized indices obey the same scaling relations as the ideal ones and the  transformations are involutions in the sense that re-renormalizing the critical exponents of the constrained system delivers their original, ideal counterparts.
Here we examine Fisher renormalization of critical amplitudes and show that, unlike for critical exponents, the associated transformations are not involutions. 
However, for ratios and combinations of amplitudes which are universal, Fisher renormalization is involutory. 

%
%-----------------------------------------------------------------------
                        \end{abstract} }
%-----------------------------------------------------------------------
%
  \thispagestyle{empty}
%
%***********************************************************************
%
  \newpage
%
%-----------------------------------------------------------------------
                  \pagenumbering{arabic}
%-----------------------------------------------------------------------

%%%%%%%%%%%%%%%%%%%%%%%%%%%%%%%%%%%%%%%%%%%%%%%%%%%%%%%%%%%%%%%%%%%
\noindent
\section{Introduction}
%%%%%%%%%%%%%%%%%%%%%%%%%%%%%%%%%%%%%%%%%%%%%%%%%%%%%%%%%%%%%%%%%%
\setcounter{equation}{0}

Universal amplitude ratios feature in all types of phase transitions and are analogous to the scaling relations which connect the various critical exponents describing power-law divergences.
{{Extensive reviews of the topic  are given in Refs.~\cite{PHA,PeVi02}, which, besides the general theory, also contain}} discussions of experimental relevance and results.
%Recent work has identified universal amplitude ratios are also of importance for  corrections to scaling \cite{IzHu01,ShJa10,Iz11,Iz12}.
One aspect that appears to be neglected in the literature is amplitude ratios in thermodynamic systems subject to constraint, a topic of importance for real systems \cite{Perk1,Perk2,Perk3}. 
The effects of such constraints on the critical exponents of experimental measurements are well known and well understood; the exponents may differ significantly from their ideal, or pure, theoretical counterparts. 
In particular, if the specific heat diverges as a power-law in the ideal system, the phase transition is manifest as a finite cusp in the real system (i.e., in the experimental realization). 
This phenomenon was explained by Fisher as being due to the effect of hidden variables \cite{Fi68}.
Fisher also established elegant relations between the exponents  of the ideal and constrained systems. 
The continued theoretical and experimental interest in universal amplitude ratios  \cite{IzHu01,DeBl04,DeHe05,DeVi10,ShJa10,Iz11,Iz12}  
and in critical phenomena in constrained systems \cite{recent2001,MrFo01,recent2007,Tr08,us}, motivates our investigation into the effects on amplitude ratios of Fisher renormaliation owing to imposition of constraints.

An attractive property of Fisher renormalization applied to critical exponents is that of involution: the  exponents which describe the ideal system are obtained from Fisher renormalization of the real exponents, just as the latter result from the former.  
In other words, applying Fisher renormalization twice delivers the identity transformation. 
Also, if the ideal exponents obey scaling relations, so too do the Fisher renormalized exponents. 
(In fact both of these properties also hold for the exponents which describe logarithmic corrections to scaling in marinal circumstances \cite{us}.)
Here we derive the  universal relations between the amplitude ratios of the Fisher renormalized, real system and the original ideal system.
We show that, in contrast to the critical exponents,  Fisher renormalization of the critical amplitudes is {\emph{not}} involutory: a double application does not deliver the original amplitudes. However, applied to universal combinations of amplitude ratios, Fisher renormalization {\emph{is}} involutory. 
We also present a new amplitude ratio involving the Lee-Yang zeros and investigate the associated involutarity.
This paper, along with Ref.~\cite{us} may therefore be considered as complementing the review of Fisher renormalization contained in Ref.~\cite{PHA}.

We follow the standard notation and write the  ideal free energy as $f(t,h)$, where $t$ is the reduced  temperature and $h$ is the  reduced external field \cite{PHA,us}.
The various thermodynamic and associated functions are defined in the usual way \cite{Fi98}, e.g., 
\begin{equation}
 m(t,h) = \frac{\partial f(t,h)}{\partial h}
 \label{IzKe01}
\end{equation}
is the magnetization.
In the frequent case of symmetry under $h \rightarrow - h$, the critical point of the ideal system $t=0$ is identified by the vanishing of $m(t,h)$.
In the absence of this symmetry one may consider the vanishing of 
$\Delta m(t,h) = m(t,h)-m(t,-h)$ in the $h \rightarrow 0$ limit instead \cite{Fi68}.
We assume the following simplified (leading) critical behaviour for the ideal system
for the specific heat, magnetization, susceptibility and correlation length:
\begin{eqnarray}
 C(t,0) & = &  A_\pm|t|^{-\alpha} \,,
\label{Ca}
\\
 m(t,0) & = & B |t|^{\beta} \quad {\mbox{for~}} t < 0\,,
\label{mb}
\\
 \chi(t,0) & = & \Gamma_\pm|t|^{-\gamma} \,, 
\label{cg}
\\
 m(0,h) & = & D h^{\frac{1}{\delta}} \,,
\label{mh}
 \\
 \xi(t,0) & = & N_\pm|t|^{-\nu} \,.
\label{xn}
\end{eqnarray}
Here, and in what follows, the subscripts $+$ and $-$ refer to amplitudes for $t>0$ and $t<0$, respectively.
Note that Eq.(\ref{Ca}) for the specific heat results from an internal energy of leading form
\begin{equation}
 e(t,0) = \pm \frac{A_\pm}{1-\alpha}|t|^{1-\alpha}.
 \label{ea}
\end{equation}
The critical correlation function is
\begin{equation}
 G(t=0,h=0;x) = \frac{\Theta}{x^{d-2+\eta}}
\,.
\label{G0}
\end{equation}
For the scaling of the Yang-Lee edge, we assume that \cite{LY}
\begin{equation}
 \theta(t) = Wt^\Delta  \quad {\mbox{for~}} t > 0\,.
\label{r1}
\end{equation}

Corresponding to the fact that $\alpha$, $\gamma$ and $\nu$ represent the critical exponents 
for {\emph{both}} $t>0$ and $t<0$, the following amplitude ratios are universal:
\begin{equation}
U_0   =  \frac{A_+}{A_-}  \,, \quad
U_2   =  \frac{\Gamma_+}{\Gamma_-}  \,, \quad
U_\xi =  \frac{N_+}{N_-} \,.
\label{ratioo}
\end{equation}
The standard scaling relations between the above critical exponents are
\begin{eqnarray}
 \alpha + d \nu  & = & 2 \,,
\label{J}
\\
\alpha + 2 \beta + \gamma & = & 2  \,,
\label{R}
\\
 (\delta - 1) \beta & = & \gamma \,,
\label{G}
\\
(2 - \eta) \nu & = & \gamma \,,
\label{F}
\\
  \beta + \gamma & = & \Delta \,,
\label{YL}
\end{eqnarray}
where the dimensionality of the system is denoted $d$.
These relations are derived in the appendix where it is shown that they correspond to the following universal ratios \cite{PHA}:
\begin{eqnarray}
  R_{\xi}  & = & A_\pm N_\pm^d\,,
  \label{Rxi}
  \\
  R_{c}    & = &  \frac{A_\pm \Gamma_\pm}{B^2}\,,
  \label{Rc}
  \\
 R_{\chi}  & = & \frac{ \Gamma_\pm B^{\delta-1}}{D^\delta}\,,
  \label{Rchi}
  \\
  Q        & = &\frac{\Theta N_\pm^{2-\eta}}{\Gamma_\pm}\,,
  \label{Q}
  \\
  Z        & = & W \left({\frac{D}{B}}\right)^\delta\,.
  \label{Z}
\end{eqnarray}
The last of these is a new universal amplitude combination not appearing previously in the literature.
(A wider set of amplitude ratios may be developed, involving more quantities on the critical isotherm \cite{PHA}. To keep the present paper compact, we focus only on the above set.) 
%For completeness, the standard derivation of these universal amplitude combinations is sumarised briefly in the appendix.

The  family $S = \left\{{\alpha, \beta, \gamma, \delta, \nu, \eta, \Delta}\right\}$ of universal critical exponents characterizes the power-law  behaviour of the specific-heat, magnetization, susceptibility,  correlation length, correlation function and Yang-Lee edge of the ideal system.
There is also a family of critical amplitudes for the ideal system, $\bar{S} = \left\{{A_{\pm}, B, \Gamma_{\pm}, D, N_{\pm}, \Theta, W}\right\}$ and 
a family of universal amplitude combinations 
$\tilde{S} = \{U_0,U_2,U_\xi,R_{\xi}, R_c, R_{\chi}, Q, Z\}$. 
These latter two families are the focus of our investigation.  
We wish to determine how they transform under Fisher renormalization and whether or not the process is involutory.

For the constrained system we write 
\begin{eqnarray}
 C_X(t,0) & = &  {A_X}_\pm|t|^{-\alpha_X}  \,,
 \label{CX}
\\
 m_X(t,0) & = & B_X |t|^{\beta_X} \quad {\mbox{for~}} t < 0\,,
 \label{mX}
\\
 \chi_X(t,0) & = & {\Gamma_X}_\pm|t|^{-\gamma_X} \,, 
 \label{chiX}
\\
 m_X(0,h) & = & D_X h^{\delta_X} \,,
 \label{mhX}
\\
 \xi_X(t,0) & = & {N_X}_\pm|t|^{-\nu_X} \,,
 \label{xiX}
\\
 G_X(0,0,x) & = & \frac{\Theta_X}{x^{d-2+\eta_X}} \,,
 \label{G00X}
\\
 \theta_X(t) & = & W_X t^{\Delta_X} \quad {\mbox{for~}} t > 0\,.
 \label{r1X}
\end{eqnarray}
Here we have assumed that the critical point of the real system is also located at $t=h=0$. We will justify this in Section~2 \cite{Fi68}.
To parallel the notation in the ideal system, we introduce
\begin{equation}
 {U_X}_0   = \frac{{A_X}_+}{{A_X}_-} \,, \quad
 {U_X}_2   =  \frac{{\Gamma_X}_+}{{\Gamma_X}_-} \,, \quad
 {U_X}_\xi  =  \frac{{N_X}_+}{{N_X}_-} \,.
\end{equation}
We also define
\begin{eqnarray}
  {R_X}_{\xi} & = & {A_X}_\pm {N_X}_\pm^d\, ,
\label{RXxi}\\
  {R_X}_{c} & = & \frac{{A_X}_\pm {\Gamma_X}_\pm}{{B_X}^2}\,,
\label{RXc}\\ 
 {R_X}_{\chi} & = & \frac{ {\Gamma_X}_\pm {B_X}^{\delta_X-1}}{{D_X}^{\delta_X}}\,,
\label{RXchi}\\
  Q_X & = & \frac{\Theta_X {N_X}_\pm^{2-\eta_X}}{{\Gamma_X}_\pm}\, ,  \\
\label{QX} 
  Z_X        & = & W_X \left({\frac{D_X}{B_X}}\right)^{\delta_X}\,.
  \label{ZX}
\label{ZX}
\end{eqnarray}
Already in Ref.~\cite{Fi68}, Fisher showed how the family of renormalized exponents  $S_X^\prime=\left\{{\alpha_X, \beta_X, \gamma_X, \delta_X, \nu_X, \eta_X}\right\}$ 
relates to the ideal exponents $S^\prime=\left\{{\alpha, \beta, \gamma, \delta, \nu, \eta}\right\}$.
That finding may be summarised as  
\begin{equation}
 S_X^\prime={\cal{F}}(S^\prime)\,,
\label{Fren}
\end{equation}
where 
\begin{equation}
\alpha_X  = \frac{-\alpha}{1-\alpha}
\,,
\label{alphaX}
\end{equation}
and
\begin{equation}
\rho_X = \frac{\rho}{1-\alpha}\,,
\label{rhoX}
\end{equation}
in which $\rho$ stands for any of the exponents $\beta$, $\gamma$ or $\nu$.
The exponent $\delta$ and the anomalous dimension $\eta$ are not renormalized:
\begin{equation}
\delta_X = \delta  \,,
 \quad 
 {\mbox{and }}
 \quad 
\eta_X = \eta\,.
\label{Fre}
\end{equation}

These formulae have  appealing properties.
Firstly, if the ideal exponents $S^\prime$ obey the scaling relations (\ref{J})--(\ref{F})  then the Fisher renormalized exponents $S_X^\prime$ obey an analogous set of relations. 
Secondly, Fisher renormalization of critical exponents is an involution
in the sense that the ideal exponents are derived  from the constrained ones in the same manner as the constrained from the ideal~\cite{Sh76}:
\begin{equation}
 S^{\prime}={\cal{F}}[{\cal{F}}(S^\prime)]\,.
\label{FF}
\end{equation}

Here we wish to investigate how the set of Fisher-renormalized real amplitudes  ${\bar{S}}_X = \left\{{{A_X}_{\pm}, B_X, {\Gamma_X}_{\pm}, D_X, {N_X}_{\pm}, \Theta_X,W_X}\right\}$ 
relates to the set of ideal amplitudes 
 $\bar{S} $.
We show that these non-universal quantities are not involutory under Fisher renormalization.
We show that to achieve involutarity, one needs universal quantities. 
Indeed, the family 
$\tilde{S}$, whose Fisher-renormalized counterpart is $\tilde{S}_X = \{{U_X}_0,{U_X}_2,{U_X}_\xi,{R_X}_{\xi}, {R_X}_c, {R_X}_{\chi}, Q_X,Z_X\}$ turns out to have the desired property:
 \begin{equation}
 {\tilde{S}}={\cal{F}}[{\cal{F}}({\tilde{S}})]\,.
\label{FFFFF}
\end{equation}
Along the way we also show that the exponent $\Delta$, characterising the scaling of the Yang-Lee edge, Fisher-renormalizes as Eq.({\ref{rhoX}) and that the involutory property (\ref{FF}) therefore applies to the full set of critical exponents $S$.
Moreover, the Fisher renormalized exponent $\Delta_X$ obeys a scaling relation analogous to Eq.(\ref{YL}), namely $\beta_X+\gamma_X=\delta_X$.

%%%%%%%%%%%%%%%%%%%%%%%%%%%%%%%%%%%%%%%%%%%%%%%%%%%%%%%%%%%%%%%%%%%
\noindent
\section{Fisher Renormalization}
%%%%%%%%%%%%%%%%%%%%%%%%%%%%%%%%%%%%%%%%%%%%%%%%%%%%%%%%%%%%%%%%%%
\setcounter{equation}{0}

{{Following Refs.\cite{Fi68,us}, we consider a system under constraint, with a hidden thermodynamic variable $x$ conjugate to a field $u$. 
The central assumption is that the singular part of the free energy of the constrained system $ f_X(t,h,u)$  is structured analogously to its ideal counterpart, so that $ f_X(t,h,u) = f[t^*(t,h,u),h^*(t,h,u)]$. (A regular background term appears additionally in Ref.\cite{Fi68}, which we omit since it has no important role here.)
The quantity  $u$ may represent a chemical potential in an Ising model of a magnet, for example and $x$ may be the density of annealed non-magnetic impurities  \cite{CaWo86}. 
The ideal transition is manifest and the ideal free energy $f(t,h)$ is recovered if $u$ is fixed at $u=0$.
The constraint is expressed as
\begin{equation}
 x(t,h,u) \equiv \frac{\partial f_X(t,h,u)}{\partial u}
 = X(t,h,u)
\,,
\label{constraint227}
\end{equation}
where $X(t,h,u)$ is assumed to be an analytic function~\cite{Fi68}.  
We assume that  $h^*(t,h,u) = h  {\cal{J}}(t,h,u)$, 
so that $h^*$, and its partial derivatives with respect to both $t$ and $u$ all vanish when $h=0$.

To identify the critical point of the real system, one  writes the  magnetization  as
\begin{equation}
  m_X(t,h,u) = \left.{\frac{\partial f_X(t,h,u)}{\partial h}}\right|_{h=0}  =   m [ t^*(t,0,u),0]{\cal{J}}(t,0,u)  \,.
\label{25}
\end{equation}
Since the critical point of the real system is given by the vanishing of the right hand side, and since ${\cal{J}}(t,0,u)$ is non-vanishing, the real critical point is 
$t^*(t,0,u) = 0$.
We write the Taylor expansion for ${\cal{J}}(t,h,u)$ about the critical point as 
$  {\cal{J}}(t,h,u) = J_0 + b_1t + \dots + c_1h + \dots + c_1(u-u_c) + \dots$, where $u_c$ is the critical value of $u$ for the real system.
The critical point is therefore marked by  ${\cal{J}}(0,0,u_c) = J_0$.

The relation between $t^*$ and $t$ comes from the constraint (\ref{constraint227}).
This will be the source of the non-trivial relationship between $t^*$ and $t$. 
Expanding  $t^*(t,0,u)$ about the critical point, $t^*(t,0,u) = a_1(u-u_c) + \dots$,
where  $u_c$ and the coefficients of the expansion are non-universal.
Therefore $  x(t,0,u) =  a_1 e(t^*,0) + \dots$, which, from Eq.(\ref{ea}), is
\begin{equation}
 x(t,0,u) = \pm a_1 \frac{A_\pm}{1-\alpha} | t^*|^{1-\alpha} + \dots
\,.
\label{19}
\end{equation}
On the other hand, and again by Taylor expansion,
\begin{equation}
 X(t,0,u)  =   X(0,0,u_c) + d_1 (u-u_c) + d_2t + \dots \,.
\label{32}
\end{equation}
Again the expansion coefficients are not universal.
Comparing with Eq.(\ref{19}), $X(0,0,u_c)$ must vanish, and
%\footnote{It is satisfying to note that the signs are correct here.}
\begin{equation}
\pm a_1 \frac{A_\pm}{1-\alpha} | t^*|^{1-\alpha} 
=
\frac{d_1}{a_1} t^* + d_2t + \dots \,.
\label{22}
\end{equation}
This is the main result of Fisher renormalization.
The first term on the right dominates the left hand side in the case that  $\alpha < 0$, so that $t^*$ and $t$ are commensurate there. However, if $\alpha >0$ the renormalization from $t$ to $t^*$ is non-trivial.
Define the non-universal quantity
\begin{equation}
 a = \left[{ \frac{d_2 (1-\alpha)}{a_1}  }\right]^{\frac{1}{1-\alpha}}\,.
\end{equation}
We then obtain
\begin{equation}
 |t^*| = a \left({ \frac{|t|}{A_\pm} }\right)^{\frac{1}{1-\alpha}}
\,.
\label{main}
\end{equation}
The interpretation of this equation is that imposing the constraint is equivalent to renormalization of the  reduced temperature in the constrained system from $t^*$. %\cite{Kr99}. 
We note that this gives
\begin{equation}
 \frac{\partial t^*}{\partial t} = \frac{a}{1-\alpha} A_\pm^{-\frac{1}{1-\alpha}}
 |t|^{\frac{\alpha}{1-\alpha}}
\,.
\label{diffmain}
\end{equation}

}}

%.......................................................................
\subsection{Thermodynamic Functions for the Constrained System}
%.......................................................................

{{We wish to determine the thermodynamic functions for the constrained systems, paying particular attention to the amplitudes.}}
Differentiating {{the constrained free energy}} with respect to $t$, 
and using  Eqs.(\ref{main}) and (\ref{diffmain}), 
\begin{equation}
e_X(t,0,u)  =  \frac{\partial f_X(t,0,u)}{\partial t} 
% = \frac{\partial f(t*,0)}{\partial  t^*}\frac{\partial t^*}{\partial t}
% + \frac{\partial f}{\partial  h^*}\frac{\partial h^*}{\partial t}
 =  e(t^*,0) \frac{\partial t^*(t,0,u)}{\partial t}
% + m(t^*,0) \frac{\partial h^*}{\partial t}\,.
 =
 \pm \frac{a^{2-\alpha}}{(1-\alpha)^2} A_\pm^{\frac{-1}{1-\alpha}} |t|^{\frac{1}{1-\alpha}}\,.
 \label{eX}
\end{equation}
The specific heat for the real system is then 
\begin{equation}
  C_X(t,0,u) = \frac{\partial e_X(t,0,u)}{\partial t} = \frac{a^{2-\alpha}}{(1-\alpha)^3}A_\pm^{\frac{-1}{1-\alpha}} |t|^{\frac{\alpha}{1-\alpha}}\,.
\end{equation}
From the form (\ref{CX}), we identify 
\begin{equation}
 \alpha_X = -\frac{\alpha}{1-\alpha}\,,
\label{alphaX}
\end{equation}
and 
\begin{equation}
 {A_X}_\pm =
  a^{1+\frac{1}{1-\alpha_X}} (1-\alpha_X)^3 A_\pm^{\alpha_X-1}
 \,.
\label{Stno}
\end{equation}
This relationship is non-universal since, besides $A_\pm$, $a$ is a non-universal constant.

The magnetization for the real system is given by Eqs.(\ref{mb}) and (\ref{25})  as
\begin{equation}
 m_X(t,0,u) = J_0 B |t^*|^{\beta}\,,
\label{mmX}
\end{equation}
for $t<0$.
From Eq.(\ref{main}), this is  Eq.(\ref{mX}) with
\begin{equation}
 \beta_X = \frac{\beta}{1-\alpha}\,,
\label{betaX}
\end{equation}
and
\begin{equation}
 B_X = J_0 a^\beta \frac{B}{A_-^{\beta_X}}
 .
 \label{45}
\end{equation}

The susceptibility for the real system is obtained by differentiating the constrained magnetisation with respect to $h$. We obtain
\begin{equation}
 \chi_X(t,0,u) = J_0^2 \chi (t^*,0) = {\Gamma_X}_\pm |t|^{-\gamma_X}\,,
\label{chiiiX}
\end{equation}
where
\begin{equation}
\gamma_X = \frac{\gamma}{1-\alpha} \,,
\label{ggammaX}
\end{equation}
and
\begin{equation}
 {\Gamma_X}_\pm =  J_0^2 a^{-\gamma}  A_\pm^{\gamma_X} \Gamma_\pm \,.
\label{ggammaX2}
\end{equation}

Along the critical isotherm $t=0$, the magnetization in field is 
\[
 m_X(0,h,u) = J_0 Dh^{\frac{1}{\delta}} + {\mathcal{O}}(h).
\]
The leading term for $\delta > 1$ is of the form (\ref{mhX}) with 
\begin{equation}
 \delta_X = \delta\,,
\label{d777}
\end{equation}
unchanged, but
\begin{equation}
 D_X = J_0^{1 + \frac{1}{\delta}} D.
 \label{d7}
\end{equation}

The correlation length  renormalizes in a similar way,
\begin{equation}
 \xi_X(t) = \xi (t^*) = N_\pm |t^*|^{-\nu} = {N_X}_\pm |t|^{-\nu_X}\,,
\label{xiX}
\end{equation}
where
\begin{equation}
\nu_X = \frac{\nu}{1-\alpha} \,, 
\label{nuX}
\end{equation}
and
\begin{equation}
 {N_X}_\pm =  a^{-\nu} A_{\pm}^{\nu_X} N_\pm .
\label{nuuX}
\end{equation}

We can consider the correlation function through  derivatives of the free energy with respect to local fields $h_1=h(x_1)$ and $h_2=h(x_2)$:
\[
 G_X(t,h,u;x) = \frac{\partial^2 f_X(t,h,u)}{\partial h_1 \partial h_2} 
              = J_0^2 \frac{\partial^2 f(t^*,h^*)}{\partial h_1^* \partial h_2^*} 
               = J_0^2  G(t*,h^*,x).
\]
Setting $t^*=t=h^*=h=0$, we obtain $G_X(0,0,u;x) = J_0^2 G(0,0,x)$ or
\begin{equation}
 \Theta_X = J_0^2 \Theta
 .
 \label{ThetaXXX}
\end{equation}

Fisher renormalization of the Yang-Lee edge for $t>0$ 
comes from {{the constrained free energy}} which gives $Z_X(t,h,u)= Z(t^*,h^*)$.
Since the edge of the distribution of zeros for the ideal system is given by $h=\theta(t) = Wt^{\Delta}$ in Eq.(\ref{r1}), the zeros' edge for the constrained system in the $h^*$-plane is $h^*=\theta(t^*) = W{t^*}^{\Delta}$.
{{Now, since $h^* = J_0h$ to leading order, }} the edge for the constrained system scales in the complex $h$-plane as
\begin{equation}
 {\theta}_X(t) = J_0^{-1} \theta(t^*) = J_0^{-1} W  {t^*}^{\Delta} = {W_X} t^{\Delta_X}\,,
\label{YLX}
\end{equation}
with $t>0$, where
\begin{equation}
\Delta_X = \frac{\Delta}{1-\alpha}  \quad \quad \quad
 {W_X} = W \frac{ a^{\Delta} 
                                        }{J_0 A_+^{\Delta_X}} \,.
\label{DeltaX}
\end{equation}

Eqs.(\ref{alphaX}), (\ref{betaX}), (\ref{ggammaX}), (\ref{d777}) and (\ref{nuX})    give the Fisher renormalization of  critical exponents, first derived in Ref.\cite{Fi68}.
Eqs.(\ref{Stno}), (\ref{45}), (\ref{ggammaX2}), (\ref{d7}) and (\ref{nuuX}) give the corresponding formula for Fisher renormalization of the critical amplitudes and are new.
Eq.(\ref{ThetaXXX}) renormalizes the amplitude of the correlation function and Eqs.(\ref{YLX}) and (\ref{DeltaX}), which govern the Yang-Lee edge, are also new results.
While Fisher renormalization of the critical exponents is involutory (meaning renormalization of renormalized exponents delivers the pure values), it is straightforward to see from the above formulae that this is not the case for the amplitudes. 
One suspects this may be because the critical exponents are universal but the critical amplitudes are not.
To investigate further, we examine Fisher renormalization of universal amplitude ratios. In the next section we show that the associated transformations are indeed involutions.

%%%%%%%%%%%%%%%%%%%%%%%%%%%%%%%%%%%%%%%%%%%%%%%%%%%%%%%%%%%%%%%%%%%
\noindent
\section{Fisher Renormalization of Universal Quantities}
\setcounter{equation}{0}
%%%%%%%%%%%%%%%%%%%%%%%%%%%%%%%%%%%%%%%%%%%%%%%%%%%%%%%%%%%%%%%%%%

We already know  that the Fisher renormalization  of the critical exponents is involutory. For example, repeated application of Eq.(\ref{alphaX}) delivers
\[
 \alpha_{XX} = - \frac{\alpha_X}{1-\alpha_X} = \alpha
\]
However, it is clear that not all quantities transform as  involutions. Considering the 
specific heat amplitudes, for example, two successive applications of (\ref{Stno}) give ${A_{XX}}_\pm$  different  from $A_\pm$.

While the individual amplitudes $A_+$ and $A_-$ are non-universal, their ratio $U_0$ is. From Eq.(\ref{Stno}), the amplitude ratio for the specific heat of the real system transforms non-trivially under Fisher renormalization as
\begin{equation}
 {U_X}_0 = \frac{{A_X}_+}{{A_X}_-} = \left({ \frac{A_+}{A_-} }\right)^{\frac{-1}{1-\alpha}} 
 = 
 U_0^{\frac{-1}{1-\alpha}} .
\label{resultA}
\end{equation}
Similarly
\begin{equation}
 {U_X}_2 = \frac{{\Gamma_X}_+}{{\Gamma_X}_-} = \frac{A_+^{\gamma_X}\Gamma_+}{A_-^{\gamma_X}\Gamma_-} = 
 U_0^{\gamma_X} U_2 \,,
\label{resultB}
\end{equation}
and
\begin{equation}
 {U_X}_\xi = \frac{{N_X}_+}{{N_X}_-} = \frac{N_+}{N_-} = 
  U_\xi\,.
\label{resultC}
\end{equation}
We observe that the transformations in these quantities between the ideal and real systems are involutory e.g., 
\begin{equation}
 {U_{XX}}_0 = {U_X}_0^{\frac{-1}{1-\alpha_X}} =  \left({{U}_0^{\frac{-1}{1-\alpha}}}\right)^{\frac{-1}{1-\alpha_X}} = U_0\,.
\label{involA}
\end{equation}

The more complex universal amplitude combinations are (\ref{Rxi}), (\ref{Rc}), (\ref{Rchi}), (\ref{Q}) and (\ref{Z}).
The non-universal terms $J_0$ and $a$ which accompany the transformations of the individual amplitudes drop out of the transformations of the universal combinations through the scaling relations (\ref{J})--(\ref{F}).
%These also transform non-trivially uner Fisher renormalization. 
These transformations are
\begin{eqnarray}
 {R_X}_c & = & \frac{1}{(1-\alpha)^3}R_c\,, 
 \label{Rctrans} \\
 {R_X}_\chi & = & R_\chi \,,
 \label{Rchitrans} \\
 {R_X}_\xi & = & \frac{1}{(1-\alpha)^3} R_\xi \,,
 \label{Rxitrans} \\
 Q_X & = & Q \,,
 \label{RQtrans} \\
 Z_X & = & \frac{Z}{U_0^{\Delta_X}} \,.
 \label{RZtrans} \\
\end{eqnarray}
Two successive applications of these transformations reveal the involutory nature of these universal combinations.

%%%%%%%%%%%%%%%%%%%%%%%%%%%%%%%%%%%%%%%%%%%%%%%%%%%%%%%%%%%%%%%%%%%
\noindent
\section{Conclusions}
\setcounter{equation}{0}
%%%%%%%%%%%%%%%%%%%%%%%%%%%%%%%%%%%%%%%%%%%%%%%%%%%%%%%%%%%%%%%%%%

We have determined how critical amplitudes transform under Fisher renormalization, a process required to determine real scaling from its ideal counterpart for systems under constraint. 
We have shown that, unlike the critical exponents, critical amplitudes do not renormalize as involutions.
We hypothesise that this is because the amplitudes, unlike the critical exponents, are non-universal.
We then showed that universal amplitude ratios are indeed involutory under Fisher renormalization.
We have also determined the Fisher renormalization of the amplitude related to the Yang-Lee zeros and showed that a related universal amplitude ratio also transforms as an involution under Fisher renormalization.

Examples of experimental systems which may be expected to manifest the phenomena described here
include magnets and fluids with specified levels of impurities \cite{Fi68}, 
e.g., Ising ferrofluids with  configurational annealed disorder \cite{Nijmeijer}, 
ternary mixtures \cite{ternary}, 
dilute polymer blends \cite{blends},
polydisperse polymeric solutions \cite{Kita},
compressible ammonium chloride \cite{old3,old4}, 
superfluidity in 3He-4He confined films \cite{DeBl04,old5}, 
nematic-smectic-A transitions in liquid-crystal mixtures \cite{old6} and emulsions \cite{old7},
and dilute antiferromagnets in applied fields \cite{CaWo86,old8}. 
Such systems are extensively discussed in the reviews \cite{PHA,PeVi02} and references therein. 
Experimental realisations of Lee-Yang zeros are also possible and discussed in Refs.\cite{Binek,Liu}.
As stated, Fisher renormalization of amplitude ratios  in these systems is a neglected topic.
It is to be hoped that the theory presented here may inspire future experimental and numerical studies in these directions.

~ \\ 
~ \\
%%%%%%%%%%%%%%%%%%%%%%%%%%%%%%%%%%%%%%%%%%%%%%%%%%%%%%%%%%%%%%%%%%%
\noindent
{\bf{Acknowledgements:}}
%%%%%%%%%%%%%%%%%%%%%%%%%%%%%%%%%%%%%%%%%%%%%%%%%%%%%%%%%%%%%%%%%%
We thank  B. Berche, J. Flanagan Jones and Yu. Holovatch for checking some of the calculations.
This work was supported by a Marie Curie International Incoming Fellowship (Project no. 300206-RAVEN) and the International Research Staff Exchange Scheme (Projects no. 295302-SPIDER and 612707-DIONICOS) 
within 7th European Community Framework Programme as well as
by the grant of the Science Committee of the Ministry of Science and Education of the Republic of Armenia under Contract 13-1C080.
This work was also partly supported by the 
Nancy-Leipzig-Coventry-Lviv Doctoral College for the Statistical Physics of Complex Systems.

%%%%%%%%%%%%%%%%%%%%%%%%%%%%%%%%%%%%%%%%%%%%%%%%%%%%%%%%%%%%%%%%%%%
\noindent
\appendix
\section{Appendix: Universal Amplitude Combinations}
\setcounter{equation}{0}
%%%%%%%%%%%%%%%%%%%%%%%%%%%%%%%%%%%%%%%%%%%%%%%%%%%%%%%%%%%%%%%%%%

We briefly remind how to identify the universal amplitude combinations, beginning with the standard scaling form for the  free energy and correlation length \cite{PHA,PeVi02,Fi98,BBS}
\begin{equation}
 f(t,h)    =   b^{-d} Y(K_t b^{y_t} t, K_h b^{y_h}h)\,, 
 \quad 
 \xi(t,h)  =   b      X(K_t b^{y_t} t, K_h b^{y_h}h)\,.
\label{A1}
\end{equation}
The scaling functions $Y$ and $X$ are universal and all the non-universality is contained in the metric factors  $K_t$ and $K_h$.
Differentiating the free energy with respect to $h$ or $t$
delivers the scaling form for the magnetization, susceptibility and specific heat by chosing 
$ b = K_t^{-{1}/{y_t}} |t|^{-{1}/{y_t}}$ or $ b = K_h^{-{1}/{y_h}} h^{-{1}/{y_h}}$ appropriately. 
One then eliminates the scaling dimensions $y_t$, $y_h$ by expressing them in terms of $\beta$ and $\delta$ (for example), and the metric factors by writing them in terms of $B$ and $D$.
The resulting expressions of  $\alpha$ and $\gamma$ in terms of $\beta$ and $\delta$ 
deliver the static scaling relations (\ref{R}) and (\ref{G}). 
Correspondingly, one can express $A_\pm$ and $\Gamma_\pm$ in terms of $B$ and $D$,
\begin{eqnarray}
 \Gamma_\pm & = & 
 \frac{Y^{(hh)}(\pm 1,0)}{
                        \left[{Y^{(h)}(1,0)}\right]^{\frac{1}{\beta}} 
                        \left[{Y^{(h)}(0,1)}\right]^{\delta} 
                        }
                        B^{1-\delta}
                        D^{\delta} \,,
 \label{A31}
 \\
 A_\pm & = & 
 \left[{Y^{(h)}(1,0)}\right]^{-(\delta + 1)}
 \left[{Y^{(h)}(0,1)}\right]^{\delta}
 Y^{(tt)}(\pm 1, 0)
 \frac{B^{\delta+1}}{D^\delta}
  \,.
 \label{32}
 \end{eqnarray}
From the first of these, $\Gamma_\pm B^{\delta-1}/D^\delta$ is a universal combination of universal factors. This is $R_\chi$ in Eq.(\ref{Rchi}).
From the second, the quantity $R_c$ in Eq.(\ref{Rc}) is seen to be universal.

From Eqs.(\ref{A1}), the correlation length is
$ \xi(t,0) = N_{\pm} |t|^{-\nu}$ where
\begin{equation}
 \nu = \frac{1}{y_t}
 \quad
 \mbox{and}
 \quad
  N_\pm = K_t^{-\frac{1}{y_t}}X(\pm 1,0).
 \label{A35}
\end{equation}
From the expression for $\alpha$ in terms of the scaling dimensions, the first of these delivers  the hyperscaling relation (\ref{J}).
To connect $N_\pm$ to the other amplitudes, one can exploit the relatonship between the susceptibility and the correlation function,
\begin{equation}
 \chi = \int_0^{\xi}{G(x) x^{d-1} dx} = \Theta \xi^{2-\eta},
\end{equation}
from which Fisher's scaling relation ({\ref{F}) follows, along with
\begin{equation}
 \Gamma_\pm = \Theta N_\pm^{2-\eta}.
 \label{A42}
\end{equation}
The combination $Q=\Theta N_{\pm}^{2-\eta}/\Gamma_{\pm}$ of Eq.(\ref{Q}) is therefore universal.
Similarly, the universality of $R_\xi$ in Eq.(\ref{Rxi}) can be explained through the hyperscaling relation 
$f(t,0) = A_\pm |t|^{2-\alpha}/(2-\alpha)(1-\alpha) \sim \xi^d(t,0)  = (N_\pm|t|^{-\nu})^d$.

Finally, chosing $ b = K_t^{-{1}/{y_t}} |t|^{-{1}/{y_t}}$  in Eq.(\ref{A1}), the partition function must take the form \cite{IPZ}
\begin{equation}
 Z(t,h) \propto Q\left({K_hK_t^{-\frac{y_h}{y_t}} |t|^{-\frac{y_h}{y_t}}h}\right).
 \label{A44}
\end{equation}
The Lee-Yang zeros are given by $Q=0$ or $h = W t^\Delta$
where 
\begin{equation}
 \Delta = \frac{y_h}{y_t}
 \quad 
 \mbox{and}
 \quad
 W = Q^{-1}(0) K_t^{\Delta} K_h^{-1},
 \label{A47}
\end{equation}
where $Q^{-1}$ is an inverse function.
The scaling relations then give $\Delta = \beta \delta = \beta + \gamma$ while the forms for the metric factors  give
\begin{equation}
 W = Q^{-1}(0) \left[{\frac{Y^{(h)}(0,1)}{Y^{(h)}(1,0)}}\right]^{\delta} \left({\frac{B}{D}}\right)^\delta.
 \label{A50}
\end{equation}
Although $W$, $B$ and $D$ are non-universal, the combination $W(D/B)^\delta$ is universal.
We denote this new amplitude ratio by $Z$ in Eq.(\ref{Z}).

\bigskip
%
%%%%%%%%%%%%%%%%%%%%%%%%%%%%%%%%%%%%%%%%%%%%%%%%%%%%%%%%%%%%%%%%%%%

\end{document}